\documentclass[10pt,conference]{IEEEtran}

\def\argmax{\mathop{\rm arg\,max}}
\def\Var{\mathop{\rm Var}}

\usepackage{times}
\usepackage{amssymb}
\usepackage{amsmath}
\usepackage{epsfig}

\newcommand{\be}{\begin{eqnarray}}
\newcommand{\ee}{\end{eqnarray}}

\begin{document}

\renewcommand{\textfraction}{0}
\newtheorem{theorem}{Theorem}
\newtheorem{lemma}{Lemma}
\newtheorem{definition}{Definition}
\newtheorem{remark}{Remark}

\title{On the Capacity of Partially\\ Cognitive Radios}

\author{
\authorblockN{G. Chung, S. Sridharan, and S. Vishwanath}
\authorblockA{Wireless Networking and Communication Group\\
University of Texas at Austin\\
Austin, TX 78712, USA\\
Email: \{gchung,sridhara,sriram\}@ece.utexas.edu}
\and
\authorblockN{C. S. Hwang}
\authorblockA{Communication Lab., SAIT\\
Samsung Electronics Co. Ltd.\\
Yongin, Korea\\
Email: cshwang@samsung.com}
}
\maketitle
\begin{abstract}
This paper considers the problem of cognitive radios with partial-message information. Here, an interference channel setting is considered where one transmitter (the ``cognitive'' one) knows the message of the other (``legitimate'' user) {\em partially}. An outer bound on the capacity region of this channel is found for the ``weak'' interference case (where the interference from the cognitive transmitter to the legitimate receiver is weak). This outer bound is shown for both the discrete-memoryless and the Gaussian channel cases. An achievable region is subsequently determined for a mixed interference Gaussian cognitive radio channel, where the interference from the legitimate transmitter to the cognitive receiver is ``strong''. It is shown that, for a class of mixed Gaussian cognitive radio channels, portions of the outer bound are achievable thus resulting in a characterization of a part of this channel's capacity region.

Note that results in this paper specialize to the case of the weak/mixed interference channel and the cognitive radio channel with full-message information.
 \footnote{This work is
supported by a grant from Samsung Advanced Institute of Technology.}

\end{abstract}
\section{Introduction}
\label{sec-introduction}

A cognitive radio is one that possesses information that allows it to tailor its transmission to maximize network throughput while meeting constraints imposed on it \cite{Mitola}. There are multiple notions of cognition in literature \cite{Mitola}, \cite{Haykin} with an increasingly popular strategy known as overlay cognition, where both the cognitive and the legitimate users transmit their own messages in the same sub-band simultaneously, as in \cite{Devroye-Tarokh}. In this setting, the cognitive transmitter has access to (limited) information about the legitimate user so as to mitigate network interference and thus increase overall throughput.

In previous work, the class of interference channels with degraded message sets has been considered \cite{Yates1}, where the cognitive user has access to the entire message of the legitimate user. Examples of this setting include \cite{Vishwanath}, where the authors determine the capacity region of this channel for both the case of ``weak'' interference and for a class of ``strong'' interference channels. However, the paper's assumption of perfect and complete message information should be relaxed in order to apply the ideas and concepts to more general classes of cognitive radio channels.

This paper considers a cognitive radio channel model where the cognitive radio is not fully cognitive of the other transmitter's message set. In this setting, the cognitive radio has access only to a portion of the message. Note that as this portion varies from nothing to everything, it includes the interference channel (IFC) in literature \cite{han}, \cite{Sato}, \cite{carleial}, and IFC with fully-degraded message set \cite{Vishwanath} as special cases. This channel is referred to as an interference channel with a partially cognitive transmitter. Note that this channel model is motivated by practical constraints, where the cognitive transmitter is only able to garner limited information about the legitimate transmitter's message.

The interference channel with a partially cognitive transmitter has already been studied in \cite{Maric}, with a specific focus on strong interference settings. This paper focuses on the weak and mixed interference settings. Specifically, we derive an outer bound on the capacity region of this channel for both the discrete memoryless and Gaussian cases when the interference from the cognitive transmitter to the legitimate receiver is ``weak''. Subsequently, we show for the Gaussian case that Gaussian distributions satisfying the constraints on the inputs/auxiliary random variables which makes the outer bound extreme exist. Finally, for a special class of mixed interference channels (where the interference from the cognitive transmitter to the legitimate receiver is ``weak'' and that from the legitimate transmitter to the cognitive receiver is ``strong''), we show that a portion of the capacity region can be characterized, i.e., a non-trivial subset of the outer bound is achievable.

This paper is organized as follows. The next section details the system model and notations used in the paper. In Section III, we describe an outer bound on the partially cognitive radio channel for the discrete memoryless case and for the Gaussian channel. In Section IV, we describe an achievable region for the Gaussian partially cognitive radio channel. In Section V, we derive channel conditions under which the achievable region is optimal. We conclude in Section IV.

\section{System Model and Preliminaries}
The notation used in this paper is based largely on that of \cite{Vishwanath}. Random variables (RVs) are denoted by capital letters, and their realizations using the corresponding lower case letters. $X_m^n$ denotes the random vector $(X_m,...,X_n)$, $X^n$ denotes the random vector $(X_1,...,X_n)$, and $X^{n\backslash m}$ denotes the random vector $(X_1,...,X_{m-1},X_{m+1},...,X_n)$. Also, for any set $S$, $\overline{S}$ denotes the convex hull of $S$, and $\widetilde{S}$ means the complementary set of $S$. Finally, the notation $X\Rightarrow Y\Rightarrow Z$ is used to denote that $X$ and $Z$ are conditionally independent given $Y$.

\subsection{Discrete Memoryless Partially Cognitive Radio Channels}
A two user interference channel as in Fig. 1 is a quintuple $(\mathcal{X}_1,\mathcal{X}_2,\mathcal{Y}_1,\mathcal{Y}_2,p)$, where $\mathcal{X}_1,\mathcal{X}_2$ are two input alphabet sets; $\mathcal{Y}_1,\mathcal{Y}_2$ are two output alphabet sets; $p(y_1,y_2|x_1,x_2)$ is a transition probability. Since we confine channel to be memoryless, the transition probability of $y_1^n,y_2^n$ given $x_1^n,x_2^n$ is\\
\begin{align*}
p(y_1^n,y_2^n|x_1^n,x_2^n)=\displaystyle\prod_{i=1}^n p(y_{1,i},y_{2,i}|x_{1,i},x_{2,i})
\end{align*}
\begin{center}
\begin{figure}
\includegraphics[width=70mm]{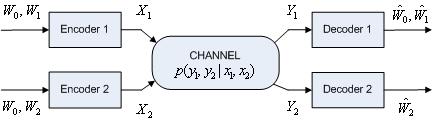}
\caption{The discrete memoryless partially cognitive radio model}
\end{figure}
\end{center}
This channel model is similar to that of an interference channel with the difference being the message sets at each transmitter. Transmitter 1 is the legitimate user, who communicates messages from the sets $W_0\in \{1,...,M_0 \}$ and $W_1\in \{1,...,M_1\}$ to Receiver 1, the legitimate receiver. Transmitter 2, the cognitive transmitter communicates messages $W_2\in \{1,...,M_2\}$ to Receiver 2, the cognitive receiver. The unique feature of this channel is that the realization of $W_0$ is known to {\em both} Transmitters 1 and 2, which  allows for partial unidirectional cooperation between the transmitters. An $(R_0,R_1,R_2,n,P_{e,0},P_{e,1},P_{e,2})$ code is any code with the rate vector $(R_0,R_1,R_2)$ and block size $n$, where $R_t\triangleq\log(M_t)/n$ bits per usage for $t=0,1,2$. As discussed above, $W_0$, and $W_1$ are the messages from Receiver 1 which must be decoded with (average) probabilities of error of at most $P_{e,0},P_{e,1}$ respectively, and $W_2$ must be retrieved at Receiver 2 while suffering an error probability of no more than $P_{e,2}$. Rate pair $(R_0,R_1,R_2)$ is said to be achievable if the error probabilities $P_{e,t}$ for $t=0,1,2$ can be made arbitrarily small as the block size $n$ grows. The capacity region of the interference channel with partially cognitive transmitter is the closure of the set of all achievable rate pairs $(R_0,R_1,R_2)$. The main goal of the users, legitimate and cognitive, is to maximize in general the ${\mu}_0 R_0+{\mu}_1 R_1 + {\mu}_2 R_2$ for some non-negative number ${\mu}_0, {\mu}_1$, and ${\mu}_2$. We also have a restriction on the pair $(R_0,R_1)$, such that $R_1  \ge \mu R_0$ for some positive number $\mu$. This restriction is to ensure that optimization of $({\mu}_0, {\mu}_1, {\mu}_2)$ in order to maximize ${\mu}_0 R_0+{\mu}_1 R_1 + {\mu}_2 R_2$ does not drive $R_1$ to zero, which results in a fully cognitive solution.
\subsection{Gaussian Partially Cognitive Radio Channel}
In the Gaussian IFC, input and output alphabets are the reals $\mathbb{R}$, and outputs are the linear combination of the inputs and additive white Gaussian noise. A Gaussian IFC model in Fig 2. is characterized mathematically as follows:
\begin{eqnarray}\label{eqn : channel}\
Y_1 &=X_1+bX_2+Z_1\nonumber\\
Y_2 &=aX_1+X_2+Z_2,
\end{eqnarray}
where $a$ and $b$ are real numbers and $Z_1$ and $Z_2$ are independent, zero-mean, unit-variance Gaussian random variables. Further, each transmitter has a  power constraint
\begin{align*}
\frac{1}{n}\displaystyle\sum_{i=1}^n{\mathbb{E}}[X_{t,i}^2]\leq P_t,  t=1,2.
\end{align*}
\begin{center}
\begin{figure}
\includegraphics[width=70mm]{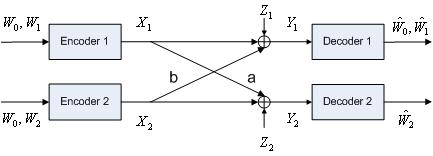}
\caption{The Gaussian partially cognitive radio channel}
\end{figure}
\end{center}
This concludes our description of the models considered in this paper. The next section describes the outer bound on the capacity region for these channels under ``weak'' interference.

\section{The Outer Bound region}
\subsection{Discrete Memoryless Partially Cognitive Radio Channels}
For a discrete memoryless channel, under the condition
\begin{align}
X_2|X_1\Rightarrow Y_1|X_1\Rightarrow Y_2|X_1,
\end{align}
we say that the legitimate receiver is observing weak interference.
For the Gaussian case, the weak interference constraint can be interpreted as the requirement of $b<1$ in (1).
First, we reproduce a useful lemma from \cite{Yates}.
\begin{lemma} [\cite{Yates}]
\label{Lemma1}
The following forms a Markov chain for the partially cognitive radio channel:
\begin{align}
(W_0,W_t)\Rightarrow (W_0,X_t)\Rightarrow Y_t
\end{align}
where $t=1,2$.
\end{lemma}
We present the outer bound in the following:
\begin{theorem}
\label{main}
The convex closure of the following inequalities defines an outer bound on the capacity region of ``weak'' partially cognitive radio channels:
\begin{align}
R_0 & \le I(U,X_{1};Y_{1}|V)\\
R_0 + R_1 & \le  I(U,X_{1};Y_{1})\\
R_2 & \le I(X_{2};Y_{2}|U,X_{1})\\
R_1 & \ge \mu R_0
\end{align}
for any $p(u,v))p(x_1|u,v)p(x_2|u)$ such that:\\
\indent
1. $V$ and $X_2$ are independent\\
\indent
2. $X_1$ is a function of $U$ and $V$\\
\indent
3. $(U,V) \Rightarrow (X_1,X_2) \Rightarrow (Y_1,Y_2)$.
\end{theorem}

\noindent{\bf Proof:} First we prove the outer bound for $R_0$ in (5) and $R_2$ in (7). We have
\begin{align*}
nR_0 & = H(W_0|W_1)\\
& \le I(W_0;Y_1^n|W_1)+n\epsilon_0\\
& = \displaystyle\sum_{i=1}^n [H(Y_{1,i}|Y_1^{i-1},W_1)-H(Y_{1,i}|Y_1^{i-1},W_0,W_1)]+n\epsilon_0\\
& \le \begin{array}{l}
\displaystyle\sum_{i=1}^n [H(Y_{1,i}|W_1)-H(Y_{1,i}|Y_1^{i-1},X_1^{n\backslash i},W_0,W_1,X_{1,i})]\\
+n\epsilon_0
\end{array}\\
& \stackrel{(a)}\le
\begin{array}{l}
\displaystyle\sum_{i=1}^n [H(Y_{1,i}|W_1)-H(Y_{1,i}|Y_2^{i-1},X_1^{n\backslash i},W_0,W_1,X_{1,i})]\\
+n\epsilon_0
\end{array}\\
& \stackrel{(b)} = \displaystyle\sum_{i=1}^n [H(Y_{1,i}|V_i)-H(Y_{1,i}|U_i,V_i,X_{1,i})]+n\epsilon_0\\
& = \displaystyle\sum_{i=1}^n I(U_i,X_{1,i};Y_{1,i}|V_i)+n\epsilon_0
\end{align*}
where $(a)$ results from the conditional Markov chain $Y_{2,i}|X_1^n\Rightarrow Y_1^{i-1}|X_1^n\Rightarrow Y_2^{i-1}|X_1^n$, which can be derived from the Markov chain for the weak interference channel, $X_2\Rightarrow Y_1\Rightarrow Y_2$, given $X_1$ in (3) as in the proof of Lemma 3.6 in \cite{Vishwanath}. $(b)$ results from identifying auxiliaries $U_i=(Y_2^{i-1},X_1^{n\backslash i},W_0)$ and $V_i={W_1}$. For $R_2$,
\begin{align*}
nR_2 & = H(W_2|W_0)\\
& \le I(W_2;Y_2^n|W_0)+n\epsilon_2\\
& \le I(W_2;Y_2^n,X_1^n|W_0)+n\epsilon_2\\
& \stackrel{(a)}=I(W_2;Y_2^n|X_1^n,W_0)+n\epsilon_2\\
& = H(Y_2^n|X_1^n,W_0)-H(Y_2^n|X_1^n,W_0,W_2)+n\epsilon_2\\
& \stackrel{(b)}\le H(Y_2^n|X_1^n,W_0)-H(Y_2^n|X_1^n,W_0,X_2^n)+n\epsilon_2\\
& \stackrel{(c)}\le \displaystyle\sum_{i=1}^n [H(Y_{2,i}|U_i,X_{1,i})-H(Y_{2,i}|U_i,X_{1,i},X_{2,i})]+n\epsilon_2\\
& = \displaystyle\sum_{i=1}^n I(X_{2,i};Y_{2,i}|U_i,X_{1,i})+n\epsilon_2
\end{align*}
where $(a)$ is due to the independence of $W_2$ and $X_1^n$, $(b)$ is from Lemma \ref{Lemma1}, and $(c)$ comes from the same definition above of $U_i={Y_2^{i-1},X_1^{n\backslash i},W_0}$.
Next, we prove the outer bound for the sum rate $R_0+R_1$ in (6). We have
\begin{align*}
n(R_0+R_1) & = H(W_0,W_1)\\
& \le I(W_0,W_1;Y_1^n)+n\epsilon_1\\
& = H(Y_1^n)-H(Y_1^n|W_0,W_1)+n\epsilon_1\\
& \stackrel{(a)}\le H(Y_1^n)-H(Y_1^n|W_0,X_1^n)+n\epsilon_1\\
& = \begin{array}{l}
\displaystyle\sum_{i=1}^n \left[
\begin{array}{l}
H(Y_{1,i}|Y_1^{i-1})\\
-H(Y_{1,i}|Y_1^{i-1},X_1^{n\backslash i},W_0,X_{1,i})
\end{array}
\right]\\
+n\epsilon_1
\end{array}\\
& \stackrel{(b)}\le \begin{array}{l}
\displaystyle\sum_{i=1}^n \left[
\begin{array}{l}
H(Y_{1,i}|Y_1^{i-1})\\
-H(Y_{1,i}|Y_2^{i-1},X_1^{n\backslash i},W_0,X_{1,i})
\end{array}
\right]\\
+n\epsilon_1
\end{array}\\
& \stackrel{(c)} \le \displaystyle\sum_{i=1}^n [H(Y_{1,i})-H(Y_{1,i}|U_i,X_{1,i})]+n\epsilon_1\\
& = \displaystyle\sum_{i=1}^n I(U_i,X_{1,i};Y_{1,i})+n\epsilon_1
\end{align*}
$(a)$ results from (Lemma \ref{Lemma1}), $(b)$ is due to the conditional Markov chain $Y_{2,i}|X_1^n\Rightarrow Y_1^{i-1}|X_1^n\Rightarrow Y_2^{i-1}|X_1^n$, and $(c)$ follows from the definition above of $U_i={Y_2^{i-1},X_1^{n\backslash i},W_0}$.
Note that the choice of auxiliary random variables automatically satisfies the constraints imposed on them in Theorem \ref{main}.
Finally, (8) comes from the restriction on the $(R_0,R_1)$, which is described in the section II.A.

\subsection{Gaussian Partially Cognitive Radio Channel}

First, note that similar proof will ensure the outer bound for the rate region defined in Theorem \ref{main} to be valid for the Gaussian partially cognitive radio channel. The main details of proof are omitted here. Next, we establish three lemmas that will be essential in proving the optimality of a jointly Gaussian input distribution for the region defined in Theorem \ref{main}.
\begin{lemma}[Lemma 1 in \cite{Thomas}]
\label{Lemma2}
Let $X_1,X_2,...,X_k$ be arbitrarily distributed zero-mean random variables with covariance matrix $K$. Let $S$ be any subset of $\{1,2,...,k\}$ and $\widetilde{S}$ be its complement. Then
\begin{align}
h(X_S|X_{\widetilde{S}})\leq h(X_S^*|X_{\widetilde{S}}^*),
\end{align}
where $X_1^*,X_2^*,...,X_k^*\sim N(0,K)$.
\end{lemma}

\begin{lemma}
\label{Lemma3}
Let $X_1,X_2,V$ be an arbitrarily distributed zero-mean random variables with covariance matrix $K$, where $X_2$ and $V$ is independent of each other. Let $X_1^*,X_2^*,V^*$ be the zero mean Gaussian distributed random variables with the same covariance matrix as $X_1,X_2,V$. Then,
\begin{align}
{\mathbb{E}}[X_1X_2] = {\mathbb{E}}[X_1^*X_2^*|V^*]
\end{align}
\end{lemma}
\noindent{\bf Proof:} Without loss of generality $X_1^*$ can be written as $X_1^* = W^* + c V^* $, where $W^*$ is the zero mean Gaussian random variable independent of $V^*$. Then
\begin{align*}
{\mathbb{E}}[X_1X_2] & = {\mathbb{E}}[X_1^*X_2^*]\\
& = {\mathbb{E}}[{\mathbb{E}}[X_1^*X_2^*|V^*]]\\
& = {\mathbb{E}}[{\mathbb{E}}[(W^* + c V^*)X_2^*|V^*]]\\
& = {\mathbb{E}}[{\mathbb{E}}[W^*X_2^*|V^*]] + c {\mathbb{E}}[{\mathbb{E}}[V^*X_2^*|V^*]]\\
& \stackrel{(a)}{=} {\mathbb{E}}[X_1^*X_2^*|V^*] + c {\mathbb{E}}[V^* {\mathbb{E}}[X_2^*]]\\
& \stackrel{(b)}{=} {\mathbb{E}}[X_1^*X_2^*|V^*]
\end{align*}
where $(a)$ results from the independence of $X_2^*$ and $V^*$. And, $(b)$ results from the fact that $X_2^*$ is zero mean.

\begin{lemma}
\label{Lemma4}
\begin{equation*}
{\mathbb{E}}[X_1^*X_2^*|V^*] \le ({\mathbb{E}}[(X_1^*)^2|V^*])^{\frac{1}{2}}({\mathbb{E}}[({\mathbb{E}}[X_2^*|X_1^*])^2])^{\frac{1}{2}}
\end{equation*}
\end{lemma}
\noindent{\bf Proof:} Note that
\begin{align*}
{\mathbb{E}}[X_1^*X_2^*|V^*] & \stackrel{(a)}{=} {\mathbb{E}}[{\mathbb{E}}[X_1^*X_2^*|V^*,X_1^*]]\\
& \stackrel{(b)}{=} {\mathbb{E}}[X_1^*{\mathbb{E}}[X_2^*|V^*,X_1^*]|V^*]\\
& \stackrel{(c)}{\le } ({\mathbb{E}}[(X_1^*)^2|V^*])^{\frac{1}{2}}({\mathbb{E}}[({\mathbb{E}}[X_2^*|V^*,X_1^*])^2])^{\frac{1}{2}}\\
& \stackrel{(d)}{\le } ({\mathbb{E}}[(X_1^*)^2|V^*])^{\frac{1}{2}}({\mathbb{E}}[({\mathbb{E}}[X_2^*|X_1^*])^2])^{\frac{1}{2}}
\end{align*}
where $(a)$ comes from the law of iterated expectations, $(b)$ from the independence of $X_2^*$ and $V^*$, $(c)$ from the Cauchy-Schwartz inequality, and $(d)$ from the fact that entropy can only be reduced by conditioning.\\

\begin{definition}
Define the rate region $\mathcal{R}_{out}^{\alpha , \beta}$ to be the convex hull of all rate pairs $(R_0,R_1,R_2)$ satisfying
\begin{equation}
\begin{array}{l}
R_0\le \frac{1}{2} \log\left( 1 + \frac{\beta P_1 + b^2 (1-\alpha) P_2 + 2b \sqrt{(\beta (1-\alpha)P_1P_2)}}{(1 + b^2 \alpha P_2)}\right)\\
R_0 + R_1\le \frac{1}{2} \log\left( 1 + \frac{P_1 + b^2 (1-\alpha) P_2 + 2b \sqrt{(\beta (1-\alpha)P_1P_2)}}{(1 + b^2 \alpha P_2)}\right)\\
R_2 \le \log (\alpha P_2 + 1)\\
R_1 \ge \mu R_0
\end{array}
\end{equation}
for some $\alpha \in [0,1]$ and $\beta \in [0,1]$
\end{definition}

\begin{definition}
Define the rate region $\mathcal{R}_{out}$ to be convex hull of the union of rate region $\mathcal{R}_{out}^{\alpha , \beta}$:
\begin{equation}
\mathcal{R}_{out}\triangleq \overline{\bigcup_{0\le \alpha,\beta \le 1} \mathcal{R}_{out}^{\alpha , \beta}}.
\end{equation}
\end{definition}

We denote $\mathcal{C}$ to be the capacity region of the Gaussian weak partially cognitive radio channel. An outer bound for $\mathcal{C}$ is obtained as follows.

\begin{theorem}
$\mathcal{R}_{out}$ is an outer bound of the capacity region for the Gaussian weak partially cognitive radio channel:
\begin{equation*}
\mathcal{C}\subset \mathcal{R}_{out}.
\end{equation*}
\end{theorem}

\noindent{\bf Proof:} We start from the rate region in Theorem \ref{main}.
\begin{align}\label{eqn : R_0}\
R_0 & \le I(U,X_1;Y_1|V) = h(Y_1|V) - h(Y_1|V,U,X_1)\nonumber\\
&=h(Y_1|V) - h(Y_1|U,X_1)\\
R_0 + R_1 & \le I(U,X_1;Y_1) = h(Y_1) - h(Y_1|U,X_1)\\
R_2 & \le I(X_2;Y_2|U,X_1) = h(Y_2|U,X_1) - h(N_2)
\end{align}
(\ref{eqn : R_0}) follows from the Markov chain, $V \Rightarrow (U,X_1) \Rightarrow Y_1$. First, we set
\begin{align}\label{eqn : setup}\
h(Y_2|U,X_1) = \frac{1}{2} \log (2 \pi e (1+\alpha P_2))
\end{align}
without loss of generality for some $\alpha \in [0,1]$. Note that
\begin{align}
Y_1=b(X_2+Z_1)+X_1+Z'\nonumber\\
h(Y_1|U,X_1)=h(b(X_2+Z_1)+Z'|U,X_1),
\end{align}
where $b < 1$ because legitimate receiver faces a weak interference, and $Z'$ is a Gaussian distributed random variable with variance $1-b^2$.
By entropy power inequality (EPI)\cite{Cover}, we have,
\begin{align*}
2^{2 h(Y_1|U,X_1)} & \ge 2^{2 h(b Y_2|U,X_1)}+2^{2 h(Z')}.\\
& = b^2 2^{2 h(Y_2|U,X_1)} + 2\pi e(1-b^2)\\
& = 2\pi e(1+b^2\alpha P_2),
\end{align*}
which yields
\begin{align}
h(Y_1|U,X_1) \ge \frac{1}{2} \log (2 \pi e (1+b^2 \alpha P_2)).
\end{align}
Next, we need to bound $h(Y_1)$ and $h(Y_1|V)$. Note that by setting $h(Y_2|U,X_1) = \frac{1}{2} \log (2 \pi e (1+\alpha P_2))$ we have the following result.
\begin{align}\label{eqn : bound1}\
h(Y_2|U,X_1) & \le h(X_2+Z_2|X_1)\nonumber\\
& \le h(X_2^*+Z_2|X_1^*)\nonumber\\
& = \frac{1}{2} \log(2\pi e(1+\Var (X_2^*|X_1^*))),
\end{align}
where $\Var (\cdot | \cdot)$ denotes the conditional covariance. Combining (\ref{eqn : setup}) with (\ref{eqn : bound1}), we obtain the bound
\begin{align} \label{eqn : var1}\
\Var (X_2^*|X_1^*) \ge \alpha P_2.
\end{align}
Also,
\begin{align}\label{eqn : var2}\
\Var (X_2^*|X_1^*) = {\mathbb{E}}[{(X_2^*)}^2]-{\mathbb{E}}[({\mathbb{E}}[X_2^*|X_1^*])^2].
\end{align}
From (\ref{eqn : var1}) and (\ref{eqn : var2}), we obtain,
\begin{align}
{\mathbb{E}}[({\mathbb{E}}[X_2^*|X_1^*])^2] \le (1-\alpha)P_2.
\end{align}
 Again, we set ${\mathbb{E}}[(X_1^*)^2|V^*] = \beta P_1$ for some $\beta \in [0,1]$ without loss of generality. Now combining Lemma \ref{Lemma3}, Lemma \ref{Lemma4}, and the above results,
\begin{align}
{\mathbb{E}}[X_1X_2] \le \sqrt{(\beta P_1)} \sqrt{(1 - \alpha) P_2}.
\end{align}
Therefore, we obtain the bound for $h(Y_1)$ as
\begin{align}
h(Y_1) & \le \frac{1}{2} \log\left(2 \pi e \left(
\begin{array}
{l}1 + \Var (X_1) + b^2 \Var (X_2)\\
+ 2b {\mathbb{E}}[X_1X_2]
\end{array}
\right)\right)\nonumber \\
& \le \frac{1}{2} \log\left(2 \pi e \left(
\begin{array}{l}
1 + P_1 + b^2 P_2\\
+ 2b \sqrt{\beta(1-\alpha) P_1 P_2}
\end{array}
\right)\right)
\end{align}
For $h(Y_1|V)$, note that $(Y_1^*,V^*)$ has the same covariance matrix as $(Y_1,V)$ if $Y_1=X_1^*+b X_2^*$. Also, $Y_1$ is a mean zero Gaussian distributed random variable. Thus,
\begin{align}
h(Y_1|V) \le & h(Y_1^*|V^*)\nonumber\\
= & h(X_1^*+b X_2^*+Z_1|V^*)\nonumber\\
= & \frac{1}{2} \log\left(2 \pi e \left(
\begin{array}{l}1 + \Var (X_1^*|V^*)\\
+b^2 \Var(X_2^*|V^*) \\
+ 2b {\mathbb{E}}[X_1^*X_2^*|V^*]
\end{array}
\right)\right) \nonumber\\
\le & \frac{1}{2} \log\left(2 \pi e \left(
\begin{array}{l}1 + \beta P_1 + b^2 P_2 \\
+ 2b \sqrt{(\beta (1-\alpha)P_1P_2)}
\end{array}
\right)\right),
\end{align}
which gives the desired outer bound for the capacity region.

\section{Achievable Region for the Gaussian Channel}\label{sec : achievable region}
In this section, we describe an achievable region for the Gaussian channel model described in (\ref{eqn : channel}). In deriving the achievable region, we combine superposition coding and dirty paper coding \cite{Costa}. The legitimate transmitter encodes messages $W_0$ and $W_1$ using Gaussian codebooks and superimposes them to form its final codeword. The cognitive transmitter allocates a portion of the power in communicating message $W_0$ to the legitimate receiver. The remaining power is used in encoding its own message $W_2$ using dirty paper coding treating the codewords (from $W_0$) as non-causally known interference. Then the following two definitions and theorem present the achievable region for the Gaussian partially cognitive radio channel.

\begin{definition}
Define the rate region $\mathcal{R}_{i}^{\alpha , \beta}$ to be the convex hull of all rate pairs $(R_0,R_1,R_2)$ satisfying
\begin{equation}
\begin{array}{l}
R_0 \leq \frac{1}{2}\log\left(1 + \frac{\beta P_1 + b^2 (1-\alpha)P_2 + 2b\sqrt{\beta(1-\alpha)P_1P_2}}{1 + b^2\alpha P_2}\right)\\
R_1 \leq \frac{1}{2}\log\left(1 + \frac{(1-\beta)P_1}{1 + \beta P_1 + b^2 P_2 + 2b\sqrt{\beta(1-\alpha)P_1P_2}}\right)\\
R_1 \leq \frac{1}{2}\log\left(1 + \frac{a^2(1-\beta)P_1}{1 + a^2 \beta P_1 + P_2 + 2a\sqrt{\beta (1-\alpha)P_1P_2}}\right)\\
R_2 \leq \frac{1}{2}\log(1 + \alpha P_2)
\end{array}
\end{equation}
for some $\alpha \in [0,1]$ and $\beta \in [0,1]$.
\end{definition}

\begin{definition}
Define the rate region $\mathcal{R}_{i}$ to be convex hull of the union of rate region $\mathcal{R}_{i}^{\alpha , \beta}$:
\begin{equation}
\mathcal{R}_{i}\triangleq \overline{\bigcup_{0\le \alpha,\beta \le 1} \mathcal{R}_{i}^{\alpha , \beta}}.
\end{equation}
\end{definition}

\begin{theorem}\label{thm : achievable region} For the Gaussian channel with partially cognitive radio as described in (\ref{eqn : channel}), the region described by
\begin{equation}\label{eqn : achievable region}
\mathcal{R}_{in} = \left\{(R_0, R_1, R_2) \in \mathcal{R}_{i} : R_1 \geq \mu R_0\right\}
\end{equation}
is achievable.
\end{theorem}

{\bf Proof:} In proving the theorem, we use an encoding strategy that combines superposition coding and dirty paper coding. We first describe the encoding strategy at the two transmitters.

\textit{Encoding Strategy at legitimate transmitter}: For every message $W_0 \in \{1, \ldots, M_0\}$, the legitimate transmitter generates a codeword $X_{10}^n(W_0)$ from the distribution $p(X_{10}^n) = \Pi_{i=1}^n p(X_{10}(i))$ and $X_{10}(i) \sim \mathcal{N}(0, \beta P_1)$ for some $0 \leq \beta \leq 1$. For every message $W_1 \in \{1, \ldots, M_1\}$, the legitimate transmitter generates a codeword $X_{11}^n(W_1)$ from the distribution $p(X_{11}^n) = \Pi_{i=1}^n p(X_{11}(i))$ and $X_{11}(i) \sim \mathcal{N}(0, (1-\beta)P_1)$. The legitimate transmitter then superimposes these codewords to form the net codeword $X_1^n$ as
\begin{displaymath}
X_1^n = X_{10}^n + X_{11}^n.
\end{displaymath}

\textit{Encoding strategy at cognitive transmitter}: The cognitive transmitter allocates a portion of its power in communicating the message $W_0$ to the legitimate receiver. For message $W_0$, the cognitive transmitter generates a codeword $X_{20}^n(W_0)$ as follows:
\begin{displaymath}
X_{20}^n(W_0) = \sqrt{\frac{(1-\alpha)P_2}{\beta P_1}} X_{10}^n(W_0).
\end{displaymath}
That is, the cognitive transmitter uses the same codeword for encoding message $W_0$ as used by the legitimate transmitter except that it is scaled to power $(1-\alpha)P_2$ for some $0 \leq \alpha \leq 1$. Next, the cognitive transmitter encodes message $W_2$ to codeword $X_{22}^n$. The codeword is generated using dirty paper coding treating $a X_{10}^n + X_{20}^n$ as non-causally known interference. A characteristic feature of Costa's dirty paper coding is that the codeword $X_{22}^n$ is independent of the interference $X_{20}^n + a X_{10}^n$, and is distributed as $p(X_{22}^n) = \Pi_{i=1}^n p(X_{22}(i))$ and $X_{22}(i) \sim \mathcal{N}(0, \alpha P_2)$. The cognitive transmitter superimposes the two codewords $X_{20}^n$ and $X_{22}^n$ to form its net codeword $X_2^n$. That is,
\begin{displaymath}
X_2^n = X_{20}^n + X_{22}^n.
\end{displaymath}

Next, we describe the decoding strategy and the rate constraints associated at the two receivers.

Decoding strategy at legitimate receiver: The legitimate receiver obtains the signal
\begin{displaymath}
Y_1^n = X_{10}^n + X_{11}^n + b X_{20}^n + b X_{22}^n + Z_1^n.
\end{displaymath}
The receiver first decodes message $W_1$ treating $X_{10}^n, X_{20}^n$ and $X_{22}^n$ as Gaussian noise. After decoding message $W_1$, the receiver decodes message $W_0$ by treating $X_{22}^n$ as Gaussian noise after canceling out $X_{11}^n$. In the first stage, the receiver can decode message $W_1$ successfully if
\begin{equation}\label{eqn : constraint 1 on R_1}
R_1 \leq \frac{1}{2}\log\left(1 + \frac{(1-\beta)P_1}{1 + \beta P_1 + b^2 P_2 + 2b\sqrt{\beta (1-\alpha) P_1P_2}}\right).
\end{equation}
Similarly, the receiver can decode message $W_0$ successfully if
\begin{equation}
R_0 \leq \frac{1}{2}\log\left(1 + \frac{\beta P_1 + b^2(1-\alpha)P_2 + 2b\sqrt{\beta (1-\alpha) P_1P_2}}{1 + b^2 \alpha P_2}\right).
\end{equation}

Decoding strategy at cognitive receiver: The cognitive receiver obtains the signal
\begin{displaymath}
Y_2^n = a X_{10}^n + a X_{11}^n + X_{20}^n + X_{22}^n + Z_2^n.
\end{displaymath}
Similar to the legitimate receiver, the cognitive receiver first decodes message $W_1$ treating $X_{10}^n, X_{20}^n$ and $X_{22}^n$ as Gaussian noise. The receiver can decode message $W_1$ successfully if
\begin{equation}\label{eqn : constraint 2 on R_1}
R_1 \leq \frac{1}{2}\log\left(1 + \frac{a^2 (1-\beta)P_1}{1 + a^2 \beta P_1 + P_2 + 2a \sqrt{\beta (1-\alpha)P_1P_2}}\right).
\end{equation}

After decoding message $W_1$, the cognitive receiver decodes message $W_2$ using Costa's dirty paper decoding. In decoding message $W_2$, the cognitive receiver sees only $Z_2^n$ as Gaussian noise. $X_{10}^n$ and $X_{20}^n$ do not appear as noise as they were canceled out at the encoder side using Costa's dirty paper coding. Hence, the receiver can decode message $W_2$ successfully if
\begin{equation}
R_2 \leq \frac{1}{2}\log(1 + \alpha P_2).
\end{equation}

Hence, the region described by $\mathcal{R}_{in}$ in (\ref{eqn : achievable region}) is achievable in the Gaussian partially cognitive radio channel. This completes the proof of Theorem \ref{thm : achievable region}.

\begin{remark} It should be noted here that the cognitive receiver first cancels the interference due to message $W_1$ before decoding message $W_2$. This places a constraint on rate $R_1$ given by (\ref{eqn : constraint 2 on R_1}). Ideally, we would want the constraint on $R_1$ given by (\ref{eqn : constraint 1 on R_1}) to be more binding than the constraint on $R_1$ given by (\ref{eqn : constraint 2 on R_1}). This is possible if
\begin{eqnarray}
& \frac{a^2 (1-\beta)P_1}{1 + a^2\beta P_1 + P_2 + 2a\sqrt{\beta(1-\alpha)P_1P_2}} \nonumber\\
& \geq \frac{(1-\beta)P_1}{1 + \beta P_1 + b^2 P_2 + 2b\sqrt{\beta (1-\alpha)P_1P_2}}.
\end{eqnarray}
\end{remark}

\section{Conditions of Optimality of Achievable Region}
In this section, we compare the achievable region and the outer bound and derive conditions when the two meet. 
We say that the achievable region described in Section \ref{sec : achievable region} is $(\mu_0, \mu_1, \mu_2)$ optimal if
\begin{eqnarray}
\max_{(R_0, R_1, R_2) \in \mathcal{R}_{in}} \mu_0 R_0 + \mu_1 R_1 + \mu_2 R_2\nonumber\\
= \max_{(R_0, R_1, R_2) \in \mathcal{R}_{out}} \mu_0 R_0 + \mu_1 R_1 + \mu_2 R_2
\end{eqnarray}

Let $(R_0^o, R_1^o, R_2^o)$ be $(\mu_0, \mu_1, \mu_2)$ optimal with respect to the outer bound. That is,
\begin{equation}
(R_0^o, R_1^o, R_2^o) = \argmax_{(R_0, R_1, R_2) \in \mathcal{R}_{out}} \mu_0 R_0 + \mu_1 R_1 + \mu_2 R_2.
\end{equation}
Let $(\alpha^o, \beta^o)$ be the optimal power splits at the two transmitters that maximizes the $(\mu_0, \mu_1, \mu_2)$ sum rate with respect to the outer bound. That is,
\begin{equation}
(\alpha^o, \beta^o) = \argmax_{0 \leq \alpha, \beta \leq 1} \max_{(R_0, R_1, R_2) \in \mathcal{R}_{out}^{\alpha, \beta}} \mu_0 R_0 + \mu_1 R_1 + \mu_2 R_2.
\end{equation}
Then, we have the following lemma.
\begin{lemma} $\beta^o = 1\ $ for all $\ (\mu_0, \mu_1, \mu_2)$.
\end{lemma}
The proof of the lemma follows from the observation that $\mathcal{R}_{out}^{\alpha,1} \supseteq \mathcal{R}_{out}^{\alpha,\beta}$ for all $0 \leq \beta \leq 1$. We next look at the conditions when the achievable region meets the outer bound.

We first consider the case $\mu_0 \geq \mu_1$. Then, we have
\begin{equation}
\begin{array}{l}
R_0^o + R_1^o = \frac{1}{2}\log\left(1 + \frac{P_1 + b^2(1-\alpha^o)P_2 + 2b\sqrt{\beta^o (1-\alpha^o)P_1P_2}}{1 + b^2\alpha^oP_2}\right),\vspace{0.1cm}\\
R_1^o = \mu R_0^o\vspace{0.1cm}\\
R_2^o = \frac{1}{2}\log(1 + \alpha^oP_2).\end{array}
\end{equation}
The conditions for optimality are then given by the following lemma.

\begin{lemma}
If the following two conditions are satisfied
\begin{eqnarray}
& \frac{a^2}{1 + a^2\beta^oP_1 + P_2 + 2a\sqrt{(1-\alpha^o)P_1P_2}} \nonumber\\
& \geq \frac{1}{1 + \beta^o P_1 + b^2 P_2 + 2b\sqrt{\beta^o (1-\alpha^o)P_1P_2}},
\end{eqnarray}
\begin{equation}
\begin{array}{l}
\log\left(1 + \frac{P_1}{1 +  b^2P_2}\right) \geq \mu\log\left(1 + \frac{b^2(1-\alpha^o)P_2}{1 + b^2\alpha^oP_2}\right)\end{array},
\end{equation}
then the achievable region is $(\mu_0, \mu_1, \mu_2)$ sum optimal for $\mu_0 \geq \mu_1$.
\end{lemma}

{\bf Proof:} The proof of the lemma is fairly simple and we briefly explain the two conditions.

The first condition comes from ensuring that constraint on $R_1$ in the achievable region due to decoding message $m_1$ at the legitimate receiver is more binding than the constraint due to decoding message $m_1$ at the cognitive receiver.

The second condition comes in ensuring that the point which maximizes the $(\mu_0, \mu_1, \mu_2)$ sum in $\mathcal{R}_{out}^{\alpha^0, 1}$ is also achievable. The main details of the proof are omitted here.

Next, we consider the case $\mu_0 < \mu_1$. In this case, $R_0^o, R_1^o$ and $R_2^o$ are given by
\begin{equation}
\begin{array}{l}
R_0^o = 0\\
R_1^o = \frac{1}{2}\log\left(1 + \frac{P_1 + b^2(1-\alpha^o)P_2 + 2b\sqrt{(1-\alpha^o)P_1P_2}}{1 + b^2\alpha^oP_2}\right)\\
R_2^o = \frac{1}{2}\log(1 + \alpha^o P_2).
\end{array}
\end{equation}

The condition of optimality when $\mu_0 < \mu_1$ is given by the following lemma.
\begin{lemma} When $\mu_0 < \mu_1$, if we have $\alpha^o = 1$, then the achievable region is $(\mu_0, \mu_1, \mu_2)$ sum optimal.
\end{lemma}
The proof of the lemma follows from the argument that if $\alpha^o = 1$, then the corresponding point $(R_0^o, R_1^o, R_2^o)$ is also achievable by substituting $\alpha = 1$ and $\beta = 0$.

\section{Conclusions}
\label{sec:conclude}
In this paper, we investigated the capacity region of interference channel with partially cognitive radios. For the general discrete memoryless IFC setting, we obtained the outer bound for the capacity region when the legitimate receiver observes the weak interference. And, for a mixed interference Gaussian channel, we showed that the portions of the outer bound can be achieved.

\section{Acknowledgment}
We thank Ivana Maric for useful discussions and comments.


\begin{thebibliography}{1}
\bibitem{Mitola}
J.Mitola, ``Cognitive Radio,'' Ph.D. dissertation, Royal Institute of Technology (KTH),Stockholm, Sweden, 2000.
\bibitem{Haykin}
S. Haykin, ¡°Cognitive radio: brain-empowered wireless communications,¡± {\it IEEE J. Sel. Areas in Commun.}, vol. 23, pp. 201-220, Feb. 2005.
\bibitem{Devroye-Tarokh}
N. Devroye, P. Mitran and V. Tarokh, ``Achievable Rates in Cognitive Rado Channels,'' {\it IEEE Trans.
Inform. Theory}, vol. 52, pp. 1813-1827, May 2006.
\bibitem{Maric}
I. Maric, A. Goldsmith, G. Kramer, S. Shamai (Shitz), ``On the Capacity of Interference Channel with a Partially-Cognitive Traansmitter,''{\it IEEE Trans. Inform. Theory}.
\bibitem{Yates}
I. Maric, R. Yates, ``The Strong Interference Channel with Common Information,'' {\it Allerton Conf. Communications}, Monticello, Il, Sep. 2005.
\bibitem{Yates1}
I. Maric, R. Yates, G. Kramer, ``The strong interference channel with unidirectional cooperation,'' presented at the Information Theory and Applications (ITA) Inaugural Workshop, Feb. 2006.
\bibitem{Vishwanath}
W. Wu, S. Vishwanath and A. Arapostathis, ``On the Capacity of Interference Channel with degraded Message Sets,''{\it IEEE Trans. Inform. Theory}.
\bibitem{han}
T. S. Han and K. Kobayashi, ``A new achievable rate region for the interference channel,''{\it IEEE Trans. Inform. Theory}, vol. 27, pp. 49-60, Jan. 1981.
\bibitem{Sato}
H. Sato, ``Two-user communication channels,'' {\it IEEE Trans. Inform. Theory}, vol. 23, pp. 295-304, May 1977.
\bibitem{carleial}
A. B. Carleial, ``Outer bounds on the capacity of interference channels,'' {\it IEEE Trans. Inform. Theory}, vol. 29, pp. 602-606, Jul. 1983.
vol. IT-24, pp. 60.70, Jan. 1978.
\bibitem{Thomas}
J. A. Thomas, ``Feedback can at most double Gaussian multiple access channel capacity,'' {\it IEEE Trans. Inform. Theory}, vol. 33, pp. 711-716, Sep. 1987.
\bibitem{Weingarten-Shamai}
H. Weingarten, Y. Steinberg and S. Shamai (Shitz), ``The Capacity
region of the Gaussian MIMO broadcast channel,'' {\it IEEE Trans. Inform. Theory}, vol. 52, pp. 3936-3964, Sep. 2006.
\bibitem{Costa}
M. Costa, ``Writing on dirty paper,'' {\it IEEE Trans. Inform. Theory}, vol. 29, pp. 439-441, May 1983.
\bibitem{Cover}
T. M Cover and J.A. Thomas, {\it Elements of information theory}, ser. Wiley Series in Telecommunications. New York: John Wiley $\&$ Sons Inc., 1991, a Wiley-Interscience Publication.
\end{thebibliography}
\end{document}